\documentclass[aps,twocolumn,prl,preprintnumbers,showpacs,amsmath,amssymb,superscriptaddress]{revtex4-1}
\usepackage[pdftex, colorlinks, citecolor=blue]{hyperref}   % this is for pdflatex pdf format
\usepackage{graphicx}
\usepackage{dcolumn}
\usepackage{threeparttable}
\usepackage{multirow}
\usepackage{booktabs}
\usepackage{txfonts}
\usepackage{xcolor}
\usepackage{bm}
\usepackage{amssymb}
\usepackage{amsmath}
\usepackage{latexsym}
\usepackage{epsfig}
\usepackage{amsbsy}
\usepackage{array}
\usepackage{tabularx}
\usepackage{esvect} %vector
\usepackage{extarrows}
\usepackage{float}
\usepackage[british]{babel}

\begin{document}

\title{On parent structures of near-ambient nitrogen-doped lutetium hydride superconductor}

\author{Mingfeng Liu}
\thanks{These authors contribute equally to this work.}
\affiliation{%
Shenyang National Laboratory for Materials Science, Institute of Metal Research, Chinese Academy of Sciences, 110016 Shenyang, China.
}%
\affiliation{%
School of Materials Science and Engineering, University of Science and Technology of China, 110016 Shenyang, China.
}%

\author{Xiangyang Liu}
\thanks{These authors contribute equally to this work.}
\affiliation{%
Shenyang National Laboratory for Materials Science, Institute of Metal Research, Chinese Academy of Sciences, 110016 Shenyang, China.
}%

\author{Jiangxu Li}
\thanks{These authors contribute equally to this work.}
\affiliation{%
Shenyang National Laboratory for Materials Science, Institute of Metal Research, Chinese Academy of Sciences, 110016 Shenyang, China.
}%

\author{Jiaxi Liu}
\affiliation{%
Shenyang National Laboratory for Materials Science, Institute of Metal Research, Chinese Academy of Sciences, 110016 Shenyang, China.
}%
\affiliation{%
School of Materials Science and Engineering, University of Science and Technology of China, 110016 Shenyang, China.
}%

\author{Yan Sun}
\affiliation{%
Shenyang National Laboratory for Materials Science, Institute of Metal Research, Chinese Academy of Sciences, 110016 Shenyang, China.
}%

\author{Xing-Qiu Chen}%
\affiliation{%
Shenyang National Laboratory for Materials Science, Institute of Metal Research, Chinese Academy of Sciences, 110016 Shenyang, China.
}%

\author{Peitao Liu}%
\email{ptliu@imr.ac.cn}
\affiliation{%
Shenyang National Laboratory for Materials Science, Institute of Metal Research, Chinese Academy of Sciences, 110016 Shenyang, China.
}%

\begin{abstract}
Recently, near-ambient superconductivity has been experimentally evidenced in
a nitrogen-doped lutetium hydride by Dasenbrock-Gammon \emph{et al.} [Nature 615, 244 (2023)], which yields a remarkable maximum $T_c$ of 294 K at just 10 kbar.
However, due to the difficulty of x-ray diffraction (XRD) in identifying light elements such as hydrogen and nitrogen,
the crystal structure of the superconductor  remains elusive, in particular for
the actual stoichiometry of hydrogen and nitrogen and their atomistic positions.
This holds even for its parent structure.
Here, we set out to address this issue
by performing a thorough density functional theory study
on the structural, electronic, dynamical, and optical properties of lutetium hydrides.
Through thermal and lattice dynamic analysis as well as XRD and superconductor color comparisons,
we unambiguously clarified that the parent structures are a mixture of dominant
LuH$_2$ phase of the CaF$_2$-type (instead of originally proposed LuH$_3$ structure of $Fm\bar{3}m$ space group)
and minor LuH phase of the NaCl-type.
\end{abstract}

\date {\today}
\maketitle

\emph{Introduction.}
Since Kamerlingh Onnes observed for the first time a superconducting transition on mercury in 1911~\cite{Hg1911},
the search for high-$T_c$ superconductors at ambient conditions has been a perpetual dream for both experimental and theoretical scientists.
During the last five years, boosts in the search have been witnessed in 
hydrogen-rich systems~\cite{doi:10.1063/5.0033232,FLORESLIVAS20201,SEMENOK2020100808, doi:10.1063/5.0065287,Lilia_2022}.
This originates from seminal intuitions of Neil Ashcroft that
room-temperature superconductors may be found in hydrogen under sufficiently high pressures
and hydrides at lower pressures by chemical precompression~\cite{PhysRevLett.21.1748,PhysRevLett.92.187002}.
This results in the discovery of many hydrogen-rich superconductors
that can achieve near room temperature superconductivity at megabar pressures,
such as H$_3$S~\cite{Drozdov2015}, H$_3$P~\cite{https://doi.org/10.48550/arxiv.1508.06224},
LaH$_{10}$~\cite{Drozdov2019,PhysRevLett.122.027001}, ThH$_{10}$~\cite{SEMENOK202036},
YH$_6$~\cite{https://doi.org/10.1002/adma.202006832}, YH$_{\rm 9\pm x}$~\cite{PhysRevLett.126.117003},
and Lu$_4$H$_{23}$~\cite{https://doi.org/10.48550/arxiv.2303.05117}.

However, the megabar pressures are still too high for practical applications.
This pushes the search to the ternary and quaternary superhydrides~\cite{PhysRevLett.123.097001,PhysRevB.102.014516,doi:10.1021/acs.jpcc.1c10976,10.3389/femat.2022.837651}.
Indeed, it has been theoretically predicted that the potassium doped Ca(BH$_4$)$_2$ can potentially be an ambient-pressure high-$T_c$ superconductor~\cite{PhysRevB.107.L060501}.
Very recently, near-ambient superconductivity has been experimentally evidenced in a nitrogen-doped lutetium hydride by Dasenbrock-Gammon \emph{et al.}~\cite{LuNH_Nature}.
This is remarkable achievement that a maximum $T_c$ of 294 K can be achieved at a much lower pressure of 10 kbar.
From an alchemical perspective, the fact of the doped lutetium hydrides being high-$T_c$ superconductors seems an natural extension
of already explored yttrium- or lanthanum-superhydrides superconductors, since they share similarity in the number of $d$ and $s$ valence electrons,
i.e., Lu ($5d^16s^2$), La ($5d^16s^2$), and Y ($4d^16s^2$).
In this superhydride, the N doping plays a role in providing additional carriers and suppressing the formation of H$^{-}$ anions (unfavourable for superconductivity) in the lattice.
Nevertheless, due to the difficulty of x-ray diffraction (XRD) in identifying light elements of hydrogen and nitrogen,
the crystal structure, the actual stoichiometry of hydrogen and nitrogen as well as their atomistic positions remain elusive.
This also holds for its parent structure.

In this work, we aim to identify the parent structures of this nitrogen-doped lutetium hydride superconductor
by density functional theory (DFT) calculations.
We found that simply comparing the simulated XRD to the experimental one is
not capable of unequivocally determining the crystal structures:
The LuH of zinc-blend type (ZB-LuH), LuH$_2$ of fluorite type (FL-LuH$_2$), and LuH$_3$ in the $Fm\bar{3}m$ space group
exhibit almost identical simulated XRDs that all mach well the main peaks of the experimental XRD.
However, $Fm\bar{3}m$-LuH$_3$ is thermodynamically and dynamically unstable.
The remaining small peaks of the experimental XRD can be well reproduced with the LuH phase of NaCl-type (RS-LuH),
which is dynamically stable both at 0 GPa and under pressures.
Furthermore, the optical calculations show that
FL-LuH$_2$ exhibits vanishing absorption of photons near pink color,
whereas RS-LuH and $Fm\bar{3}m$-LuH$_3$ show large absorption of photons near pink color.
Considering the experimentally observed pink color in the N-doped lutetium hydride superconductor samples~\cite{LuNH_Nature},
we therefore conclude that the dominant phase of the parent structures is most likely to be FL-LuH$_2$.

\begin{figure*}
\begin{center}
\includegraphics[width=0.90\textwidth, clip]{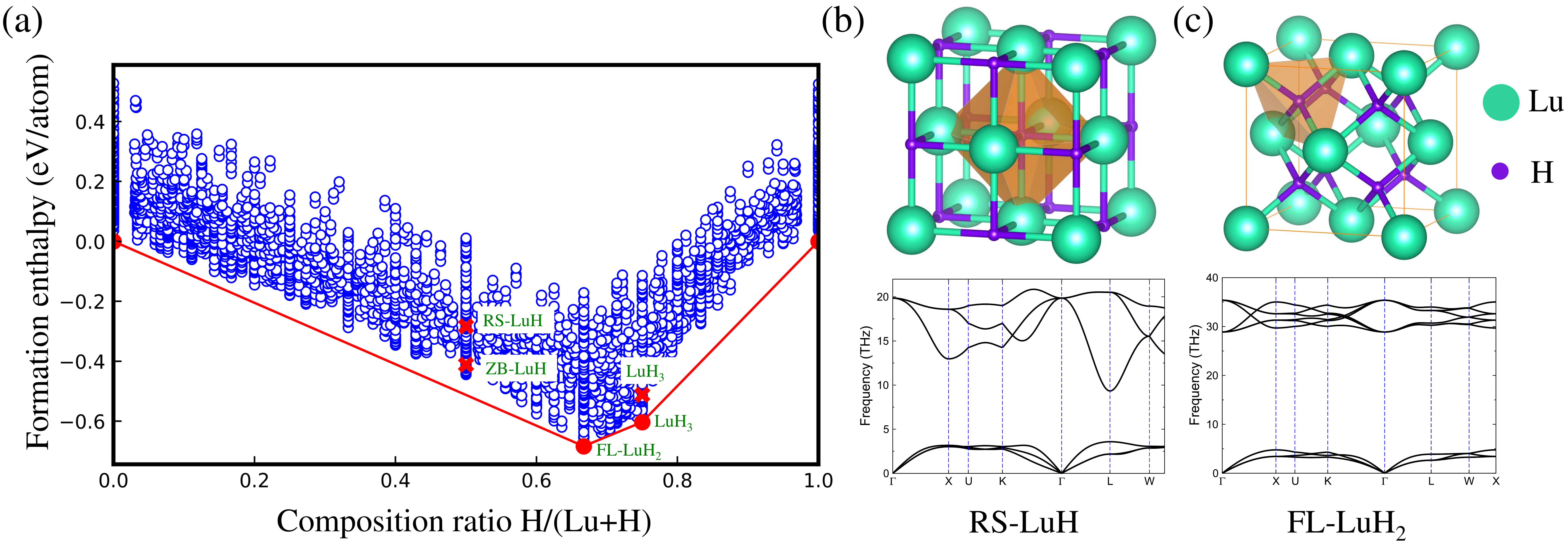}
\end{center}
\caption{
(a) Calculated formation enthalpies at 1 GPa of all the structures in the Lu-H system.
Each blue circle represents individual DFT computations
and the structures with the lowest formation enthalpies (in red solid circles) form a convex hull.
The thermodynamical instabilities of  ZB-LuH, RS-LuH, and $Fm\bar{3}m$-LuH$_3$ are indicated by red crosses.
The crystal structures and phonon dispersion relationships of (b) RS-LuH and (c) FL-LuH$_2$.
The polyhedral highlights the local environment of H.
The detailed information on other compounds is given in Supplemental Material Fig.~S1~\cite{SM}.
}
\label{Fig1_USPEC_phonon}
\end{figure*}

\emph{Computational details.}
The first-principles calculations were performed
using the Vienna \emph{ab initio} simulation package (VASP)~\cite{PhysRevB.47.558, PhysRevB.54.11169}.
The plane-wave cutoff for the orbitals was chosen to be 500 eV.
A $\Gamma$-centered $k$-point grid with a spacing of 0.03 $2\pi/\AA$
between $k$ points was used to sample the Brillouin zone.
The electronic interactions were described using the Perdew-Burke-Ernzerhof (PBE) functional~\cite{PhysRevLett.77.3865}.
The projector augmented wave (PAW) pseudopotentials~\cite{PhysRevB.50.17953,PhysRevB.59.1758}
with the valence electron configurations of $5p^{6}5d^{1}6s^2$ and $1s^1$ were employed for Lu and H, respectively.
The Gaussian smearing method with a smearing width of 0.05 eV was used.
The convergence criteria for the total energy and ionic forces were set to 10$^{-6}$ eV and 1 meV/$\AA$, respectively.
The variable-composition structure search was carried out using the generic evolutionary algorithm as implemented in the USPEX code~\cite{Oganov2006,Oganov2013},
with the maximum number of atoms being set to 32.
The phonon dispersion relationships were computed using a $3\times3\times3$ supercell
and the Phonopy code~\cite{Togo2015} based on density functional perturbation theory.
The optical absorption spectra were obtained within the independent-particle approximation
using a dense $k$-point grids with a spacing of 0.01 $2\pi/\AA$.
Note that here the Drude-like contributions stemming from intraband transitions were not considered.

\emph{Structural phase diagram and phonon dispersions.}
Figure~\ref{Fig1_USPEC_phonon}(a) shows the calculated formation enthalpies of predicted Lu-H compounds at 1 GPa
derived by the variable-composition evolutionary algorithm.
The structures with the lowest formation enthalpies forming a convex hull are thought to be the ground-state structures.
It is evident that besides the experimentally observed FL-LuH$_2$ phase~\cite{Bonnet1977},
a new predicted $R32$-LuH$_3$ structure with six Lu and eighteen H atoms in the unit cell is also on the convex hull.
It is interesting to note that $R32$-LuH$_3$ is an insulating phase 
with an indirect gap of 0.86 and 0.82 eV at 0 GPa and 1 GPa, respectively (see Supplemental Material Figs.~S1 and S2~\cite{SM}).
The $Fm\bar{3}m$-LuH$_3$ is, however, above the convex hull by 92 meV/atom, indicating its thermodynamical instability.
The ZB-LuH and RS-LuH are above the convex hull as well,
with the energies  above the convex hull being 101 and 229 meV/atom, respectively.
The phonon calculations demonstrate that ZB-LuH, RS-LuH, FL-LuH$_2$,
and $R32$-LuH$_3$ are all dynamically stable, while $Fm\bar{3}m$-LuH$_3$ exhibits soft phonon modes over the full Brillouin zone
[see Fig.~\ref{Fig1_USPEC_phonon}(c) and also Supplemental Material Fig.~S1~\cite{SM}].
We note that this is in contrast to the calculated results of Dasenbrock-Gammon \emph{et al.}~\cite{LuNH_Nature},
which, however, showed the presence of soft modes in RS-LuH.
We argue that their incorrect predictions likely originated from the employed small supercell.
For instance, using a smaller $2\times2\times2$ supercell for RS-LuH, we also reproduced the soft modes.
All the detailed information on space groups, formation enthalpies, optimized lattice parameters, and the Wyckoff positions
of predicted Lu-H binary compounds is given in Supplemental Material Table~S1~\cite{SM}.
As compared to FL-LuH$_2$, an additional hydrogen atom appears
in a hollow octahedral site of $Fm\bar{3}m$-LuH$_3$ (see Supplemental Material Fig.~S1~\cite{SM}), implying strong anharmonicity.
Therefore, the dynamical stability of $Fm\bar{3}m$-LuH$_3$ should be further carefully examined
with beyond harmonic approximation.

\begin{figure}
\begin{center}
\includegraphics[width=0.48\textwidth, clip]{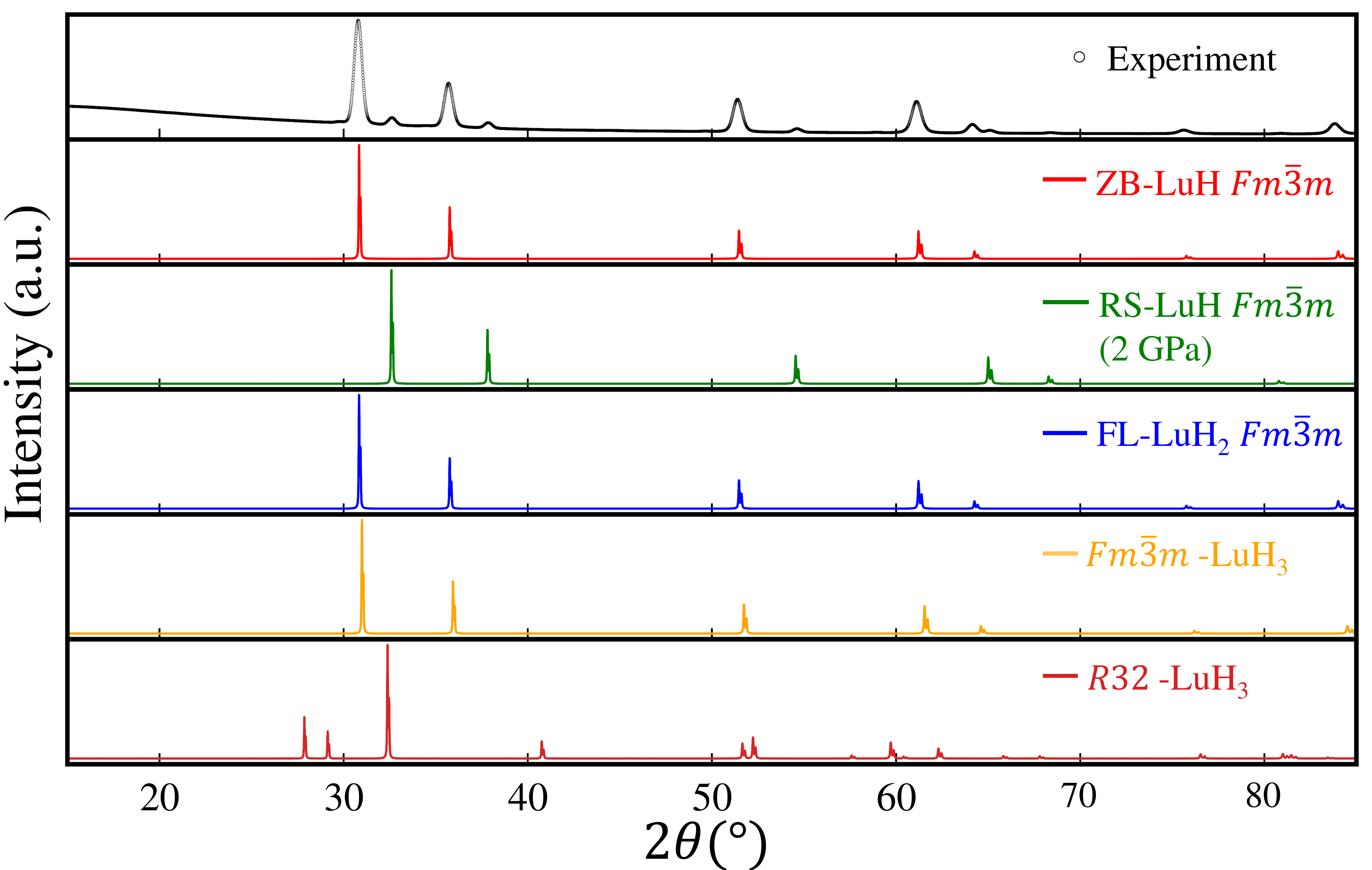}
\end{center}
\caption{Simulated XRDs of different Lu-H binary compounds at 0 GPa (except for RS-LuH,
which was obtained at 2 GPa) and compared to the experimental one obtained at 0 GPa.
The experimental data are taken from Ref.~\cite{LuNH_Nature}.}
\label{Fig2_XRD}
\end{figure}

\emph{XRD simulations.}
Figure~\ref{Fig2_XRD} compares the simulated XRDs of different Lu-H binary compounds to the experimental one obtained at 0 GPa.
One can see that ZB-LuH, FL-LuH$_2$ and $Fm\bar{3}m$-LuH$_3$
exhibit almost indistinguishable simulated XRDs that all mach well the main peaks of the experimental one.
This is not unexpected, because all of them show the similar lattice constants ($\sim$5.0~$\AA$)
(see Supplemental Material Tables~S1 and S2~\cite{SM}) and the same framework of Lu (see Supplemental Material Fig.~S1~\cite{SM}), while the subtle differences in
the number and positions of H atoms are not capable to be captured by the XRD technique.
Although the $R32$-LuH$_3$ phase is the thermodynamical ground state (on the convex hull),
its simulated XRD obviously deviates the experimental one, excluding its existence in the N-doped lutetium hydride superconductor.
The remaining small peaks appearing in the experimental XRD can be well captured by the RS-LuH phase,
indicating the existence of a small portion of the RS-LuH phase in the superconductor.
We note that the perfect agreement on the small peaks of the experimental data at 0 GPa can only be achieved by the RS-LuH phase at 2 GPa.
Reducing the pressure results in increasing lattice constants, which lead to a rigid shift toward smaller diffraction angles.
One might think that this is a result of the overestimation of lattice constants by the PBE functional.
However, if this was the reason, the presently perfect agreement on the major peaks of the experimental data
achieved by ZB-LuH, FL-LuH$_2$ and $Fm\bar{3}m$-LuH$_3$ would be somewhat deteriorated.

\begin{figure}
\begin{center}
\includegraphics[width=0.45\textwidth, clip]{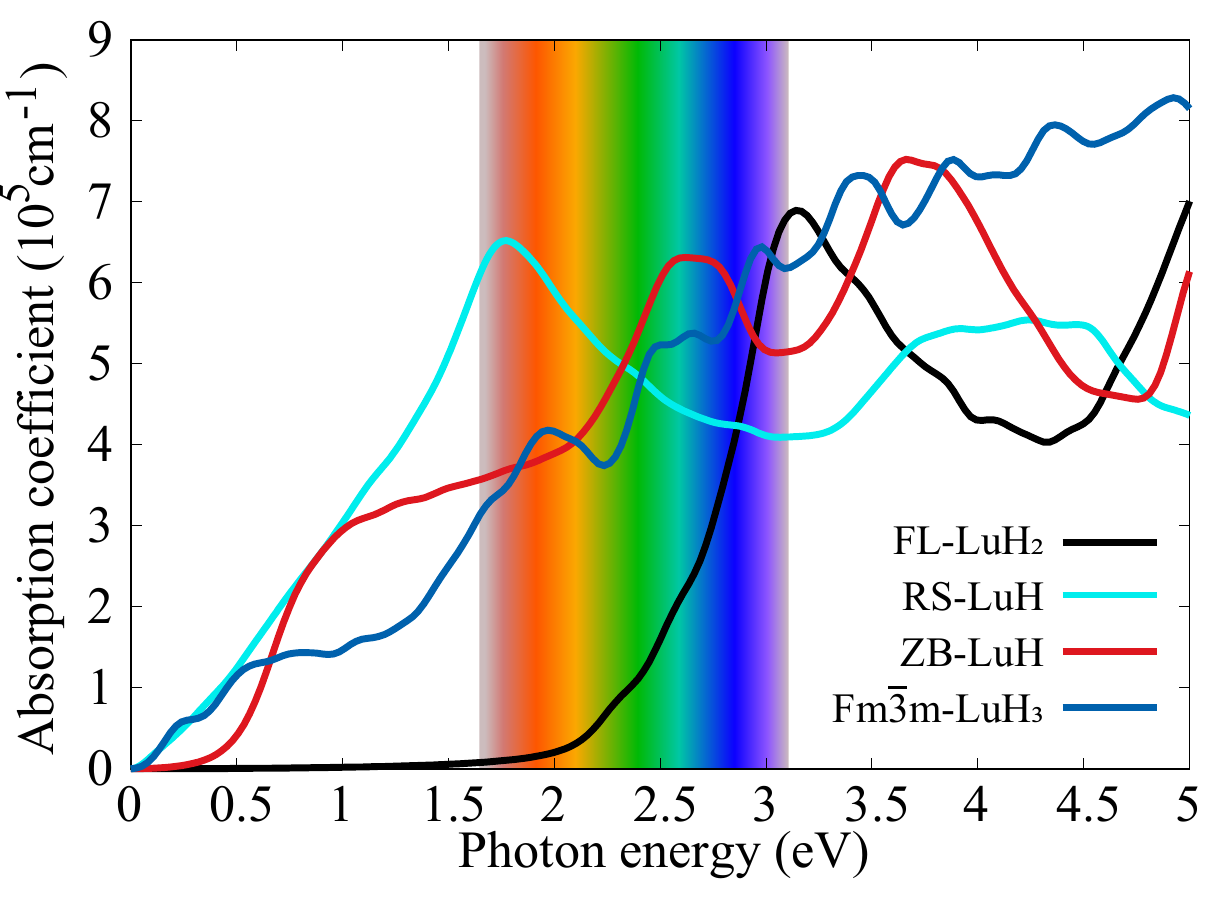}
\end{center}
\caption{Calculated optical absorption spectra of ZB-LuH, RS-LuH,
FL-LuH$_2$, and $Fm\bar{3}m$-LuH$_3$ compounds at 1 GPa.
}
\label{Fig3_optics}
\end{figure}

\emph{Optical spectra.}
Figure~\ref{Fig3_optics} displays the calculated optical absorption spectra of ZB-LuH, RS-LuH,
FL-LuH$_2$, and $Fm\bar{3}m$-LuH$_3$ compounds at 1 GPa within the independent-particle approximation.
It is evident that only FL-LuH$_2$ exhibits vanishing absorption coefficients of photons near pink color,
in good agreement with the experimental spectra~\cite{PhysRevB.19.4855}.
This arises from the existence of large direct band gaps ($\sim$1.9 eV) between the bands close to the Fermi level over the full Brillouin zone (see Supplemental  Material Fig.~S2~\cite{SM}).
By contrast, the $Fm\bar{3}m$-LuH$_3$ structure shows strong absorption of the photons near pink color.
Considering the experimentally observed pink color in the N-doped lutetium hydride superconductor samples~\cite{LuNH_Nature},
the dominant phase of the parent structures is most likely to be FL-LuH$_2$,
rather than the originally proposed $Fm\bar{3}m$-LuH$_3$ structure~\cite{LuNH_Nature}.
We note that the overall optical absorption spectra change only slightly with the pressure from 0 to 1 GPa (compare Supplemental  Material Fig.~S1 and Fig.~S2~\cite{SM}).
Our work provides a new route to identify the parent structures via the optical spectroscopy.

\begin{table}
\caption{A summary of energy above the convex hull ($E_{\rm hull}$, in meV/atom), phonon stability,
and absorption of pink-color photons as well as matching degree on the experimental XRD
for different Lu-H binary compounds at 1 GPa.}
\begin{ruledtabular}
\begin{tabular}{ccccccccc}
Phase  & $E_{\rm hull}$  & Phonon   & Absorption of        & XRD \\
          &                         &  stability  &  pink-color photons  & match\\
\hline
ZB-LuH   & 101  &  \textbf{yes}   & yes  & \textbf{dominant peaks} \\
RS-LuH  &  229 & \textbf{yes}   &  yes &  \textbf{minor peaks} \\
RL-LuH$_2$ & \textbf{0}  & \textbf{yes}   & \textbf{no} & \textbf{dominant peaks} \\
$Fm\bar{3}m$-LuH$_3$  &  92 & no   & yes  & \textbf{dominant peaks} \\
$R32$-LuH$_3$  &   \textbf{0}  & \textbf{yes}   & \textbf{no} & no match\\
\end{tabular}
\end{ruledtabular}
\label{Table1}
\end{table}

\emph{Conclusions.}
In summary, we have carried out a thorough density functional theory study
on the crystal structures, thermodynamic and dynamical stabilities, as well as optical absorption spectra of lutetium hydrides.
The results are summarized in Table~\ref{Table1}.
It is obvious that the FL-LuH$_2$ phase is the only one that satisfies all the necessary requirements toward the experimental observations,
i.e., thermodynamically and dynamically stable, no absorption of pink-color photons, and matching well the main peaks of the experimental XRD.
Therefore, we can conclude that the FL-LuH$_2$ phase is the dominant phase of the parent structures of
the recently experimentally synthesized near-ambient high-$T_c$ lutetium hydride superconductor.
The remaining small peaks of the experimental XRD can better be described by the
existence of a small portion of the RS-LuH phase, which is phonon stable at 0 GPa and under pressures.
This is further confirmed by the fact that the predicted lattice constants for FL-LuH$_2$ (5.017~$\AA$) and RS-LuH (4.800~$\AA$) at 0 GPa
agree well with the experimental values (5.029 and 4.753~$\AA$ for the so-called compound A and compound B, respectively)~\cite{LuNH_Nature}.
Given the unambiguous identification of the parent structures,
it is time to study the role of nitrogen doping in this near-ambient high-$T_c$ lutetium hydride superconductor,
since it was interestingly predicted by electron-phonon calculations that LuH$_2$ is non-superconducting at 0 GPa~\cite{LuNH_Nature}.

\emph{Acknowledgments.}
This work was supported by the National Key R\&D Program of China (No.~2021YFB3501503),
the National Science Fund for Distinguished Young Scholars (No. 51725103), 
and Chinese Academy of Sciences (No. ZDRW-CN-2021-2-5).
All calculations were performed on the high performance computational cluster at the Shenyang National University Science and Technology Park.

\bibliographystyle{apsrev4-1}
\bibliography{Reference} %here in windows there is no surfix .bib

\end{document}

\title{Supplementary Material to ``On parent structures of near-ambient nitrogen-doped lutetium hydride superconductor"}

\author{Mingfeng Liu}
\thanks{These authors contribute equally to this work.}
\affiliation{%
Shenyang National Laboratory for Materials Science, Institute of Metal Research, Chinese Academy of Sciences, 110016 Shenyang, China.
}%
\affiliation{%
School of Materials Science and Engineering, University of Science and Technology of China, 110016 Shenyang, China.
}%

\author{Xiangyang Liu}
\thanks{These authors contribute equally to this work.}
\affiliation{%
Shenyang National Laboratory for Materials Science, Institute of Metal Research, Chinese Academy of Sciences, 110016 Shenyang, China.
}%

\author{Jiangxu Li}
\thanks{These authors contribute equally to this work.}
\affiliation{%
Shenyang National Laboratory for Materials Science, Institute of Metal Research, Chinese Academy of Sciences, 110016 Shenyang, China.
}%

\author{Jiaxi Liu}
\affiliation{%
Shenyang National Laboratory for Materials Science, Institute of Metal Research, Chinese Academy of Sciences, 110016 Shenyang, China.
}%
\affiliation{%
School of Materials Science and Engineering, University of Science and Technology of China, 110016 Shenyang, China.
}%

\author{Yan Sun}
\affiliation{%
Shenyang National Laboratory for Materials Science, Institute of Metal Research, Chinese Academy of Sciences, 110016 Shenyang, China.
}%

\author{Xing-Qiu Chen}%
\affiliation{%
Shenyang National Laboratory for Materials Science, Institute of Metal Research, Chinese Academy of Sciences, 110016 Shenyang, China.
}%

\author{Peitao Liu}%
\email{ptliu@imr.ac.cn}
\affiliation{%
Shenyang National Laboratory for Materials Science, Institute of Metal Research, Chinese Academy of Sciences, 110016 Shenyang, China.
}%
	
\date {\today}
\maketitle

	\begin{table*}
		\caption{A summary of space groups, formation enthalpies (eV/atom),  optimized lattice parameters ($\AA$) and the Wyckoff positions of the Lu-H binary compounds at 0 GPa.
			The experimental lattice parameter for FL-LuH$_2$ is taken from Bonnet and Daou, Journal of Applied Physics 48, 964 (1977).}
		\begin{ruledtabular}
			\begin{tabular}{cccccccc}
				Phase          &Space       &Formation   &Lattice         & & \multicolumn{3}{c}{Atomic sites}   \\
			  &group       &energy      &parameters      &Atom    &x      &y      &z      \\
				\hline
				\specialrule{0em}{2pt}{2pt}
				ZB-LuH         &F$\bar4$3m  &$-$0.405   &a=5.0256    &Lu(4d)  &0.75      &0.75    &0.75    \\
				&           &        &        &H(4a) &0.00      &0.00   &0.00 \\
				\specialrule{0em}{2pt}{2pt}
                RS-LuH         &Fm$\bar3$m  &$-$0.277   &a=4.8002    &Lu(4a)  &0.00      &0.00    &0.00    \\
                &           &        &           &H(4b) &0.50      &0.50   &0.50 \\
				\specialrule{0em}{2pt}{2pt}
				FL-LuH$_2$     &Fm$\bar3$m     &$-$0.676   &a=5.0173    &Lu(4b)  &0.50  &0.50 &0.50 \\
				&           &        & a=5.033 (Expt.)   &H(8c)  &0.25      &0.25  &0.25      \\
				\specialrule{0em}{2pt}{2pt}
				LuH$_3$    &Fm$\bar3$m  &$-$0.503 &a=4.9930    &Lu(4a) &0.00 &0.00 &0.00     \\
				&          &       &           &H(4b) &0.50   &0.50   &0.50      \\
				&          &       &           &H(8c) &0.25    &0.25  &0.25      \\
			    \specialrule{0em}{2pt}{2pt}
			    LuH$_3$    &R32  &$-$0.506     &a=b=c=7.310    &Lu(3e) &0.8380 &0.50 &0.1620     \\
			    &          &       & $\alpha=\beta=\gamma=49.532^o$          &H(3d) &0.6620   &0.00   &0.3380      \\
			    &          &       &           &Lu(6f) &0.5910    &0.8730  &0.1970  \\
			    &          &       &           & Lu(6f)          & 0.4203      & 0.0197      & 0.7196  \\
		        &          &       &           &  H(2c)          & 0.8119     &0.8119       &   0.8119         \\
			    &          &       &           &  H(2c)         &0.3568       & 0.3568      & 0.3568            \\
			    &          &       &           &  H(1b)         & 0.50      & 0.50      & 0.50            \\
			    &          &       &           &  H(1a)         & 0.00      &  0.00     & 0.00            \\
			\end{tabular}
		\end{ruledtabular}
		\label{table1}
	\end{table*}

	\begin{table*}
	\caption{A summary of space groups, formation enthalpies (eV/atom),  optimized lattice parameters ($\AA$) and the Wyckoff positions of the Lu-H binary compounds at 1GPa.}
	\begin{ruledtabular}
		\begin{tabular}{cccccccc}
			Phase          &Space       &Formation   &Lattice         & & \multicolumn{3}{c}{Atomic sites}   \\
			&group       &energy      &parameters      &Atom    &x      &y      &z      \\
			\hline
			\specialrule{0em}{2pt}{2pt}
			ZB-LuH         &F$\bar4$3m  &$-$0.411   &a=5.0015    &Lu(4d)  &0.75      &0.75    &0.75    \\
			&           &        &           &H(4a) &0.00      &0.00   &0.00 \\
			\specialrule{0em}{2pt}{2pt}
			RS-LuH         &Fm$\bar3$m  &$-$0.283   &a=4.7769    &Lu(4a)  &0.00      &0.00    &0.00    \\
			&           &        &           &H(4b) &0.50      &0.50   &0.50 \\
			\specialrule{0em}{2pt}{2pt}
			FL-LuH$_2$     &Fm$\bar3$m     &$-$0.683   &a=5.0000    &Lu(4b)  &0.50  &0.50 &0.50 \\
			&           &        &   &H(8c)  &0.25      &0.25  &0.25      \\
			\specialrule{0em}{2pt}{2pt}
			LuH$_3$    &Fm$\bar3$m  &$-$0.511 &a=4.9779    &Lu(4a) &0.00 &0.00 &0.00     \\
			&          &       &           &H(4b) &0.50   &0.50   &0.50      \\
			&          &       &           &H(8c) &0.25    &0.25  &0.25      \\
			\specialrule{0em}{2pt}{2pt}
			LuH$_3$    &R32  &$-$0.603     &a=b=c=7.280    &Lu(3e) &0.8378 &0.50 &0.1622     \\
			&          &       & $\alpha=\beta=\gamma=49.565^o$          &H(3d) &0.6621   &0.00   &0.3379      \\
			&          &       &           &Lu(6f) &0.5904    &0.8731  &0.1973  \\
			&          &       &           & Lu(6f)          & 0.4195      &0.0201      & 0.7198  \\
			&          &       &           &  H(2c)          & 0.8115     &0.8115       &   0.8115         \\
			&          &       &           &  H(2c)         &0.3571       &0.3571      & 0.3571            \\
			&          &       &           &  H(1b)         & 0.50      & 0.50      & 0.50            \\
			&          &       &           &  H(1a)         & 0.00      &  0.00     & 0.00            \\
		\end{tabular}
	\end{ruledtabular}
	\label{table}
\end{table*}

\begin{figure}
\begin{center}
\includegraphics[width=0.80\textwidth, clip]{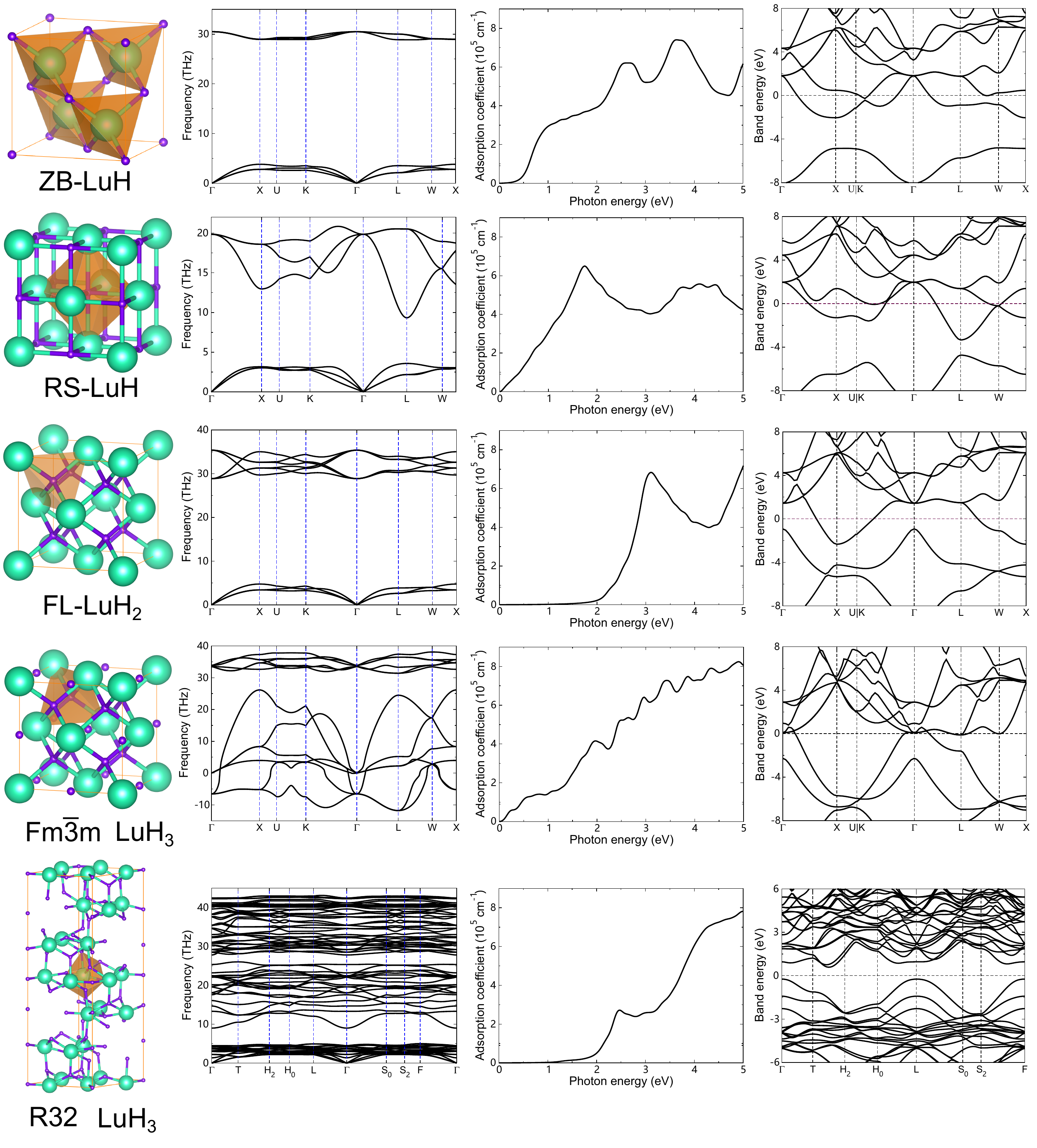}
\end{center}
\caption{A summary of crystal structures, phonon dispersion relationships, electronic band structures, and optical absorption spectra of Lu-H binary compounds at 0 GPa.
The polyhedral highlights the local environment of H.
}
\label{FigS1_0GPa}
\end{figure}

\begin{figure}
\begin{center}
\includegraphics[width=0.80\textwidth, clip]{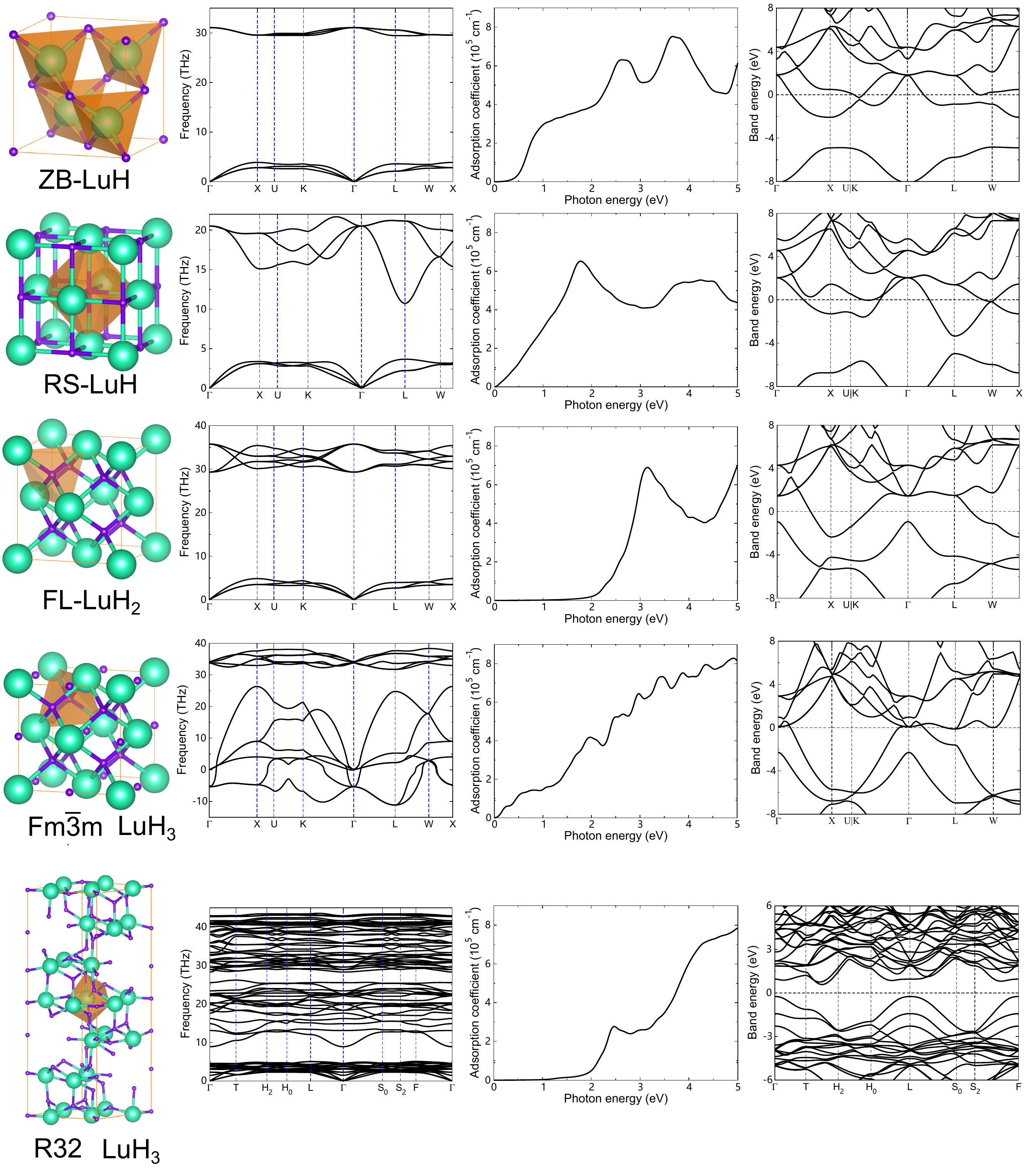}
\end{center}
\caption{Same as Fig.~\ref{FigS1_0GPa} but at 1 GPa.
}
\label{FigS2_1GPa}
\end{figure}

%\newpage
%\bibliographystyle{apsrev4-1}
%\bibliography{Reference} %here in windows there is no surfix .bib